# Nanoscale insights on the origin of the Power MOSFETs breakdown after extremely long high temperature reverse bias stress

P. Fiorenza[1,a,*], M. Alessandrino[2,b], B. Carbone[2,c], C. Di Martino[2,d], A. Russo[2,e], M. Saggio[2,f], C. Venuto[2,g], E. Zanetti[2,h], C. Bongiorno[1,i], F. Giannazzo[1,j], F. Roccaforte[1,k]

[1]Consiglio Nazionale delle Ricerche – Istituto per la Microelettronica e Microsistemi (CNR-IMM), Strada VIII, n.5 Zona Industriale, 95121 Catania, Italy

[2]STMicroelectronics, Stradale Primosole 50, 95121 Catania, Italy

[a]patrick.fiorenza@imm.cnr.it; [b]santi.alessandrino@st.com; [c]beatrice.carbone@st.com; [d]clarice.dimartino@st.com; [e]alfio-lip.russo@st.com; [f]mario.saggio@st.com; [g]carlo.venuto@st.com; [h]edoardo.zanetti@st.com; [i]corrado.bongiorno@imm.cnr.it; [j]filippo.giannazzo@imm.cnr.it; [k]fabrizio.roccaforte@imm.cnr.it

**Keywords:** Dielectric breakdown, Threading dislocation, MOSFET, HTRB

**Abstract.** In this work, the origin of the dielectric breakdown of 4H-SiC power MOSFETs was studied at the nanoscale, analyzing devices that failed after extremely long (three months) of high temperature reverse bias (HTRB) stress. A one-to-one correspondence between the location of the breakdown event and a threading dislocation propagating through the epitaxial layer was found. Scanning probe microscopy (SPM) revealed the conductive nature of the threading dislocation and a local modification of the minority carriers concentration. Basing on these results, the role of the threading dislocation on the failure of 4H-SiC MOSFETs could be clarified.

**Introduction**

4H-SiC metal oxide semiconductor field effect transistors (MOSFETs) are stepping in the automotive market for the fabrication of power components of the electric vehicles [1,2]. In this context, the reliability of 4H-SiC MOSFETs is a crucial concern to meet the safety requirements of the automotive industry. For that reason, several issues related to 4H-SiC MOSFET reliability are under discussion in the SiC community to improve the devices performances. In particular, significant efforts are focused on the stability in long-term interdiction under high temperature reverse bias stress (HTRB) [3,4].

Up to now, 4H-SiC MOSFETs are fabricated using $SiO_2$ as a gate insulator, but the breakdown kinetics of the $SiO_2$/4H-SiC system is different than the $SiO_2$/Si system. As an example, while the behavior of thin thermally grown oxides onto 4H-SiC is ideal [5,6], thicker thermal oxides (>10 nm) deviates from the ideal percolation theory, probably due to the presence of carbon-related defects in the oxide [7,8]. Moreover, premature breakdown in $SiO_2$/4H-SiC system has been also attributed to the presence of crystalline defects (pipes, carrots, bars, etc) [9], and to the step-bunching on the 4H-SiC surface [10,11].

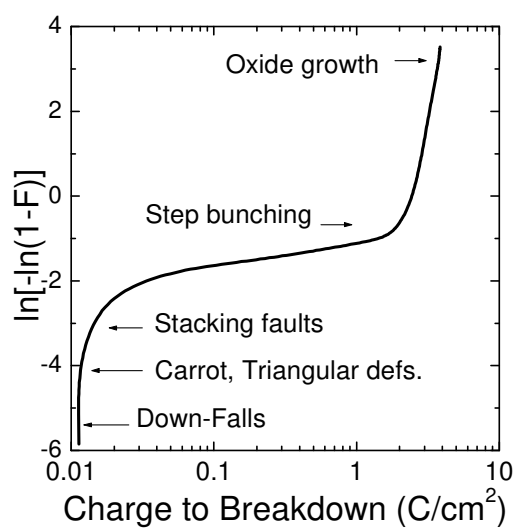

Fig. 1: Schematic representation of a typical Weibull distribution of 4H-SiC MOS capacitors referenced with defects that may affect the device lifetime.

Fig. 1 graphically depicts the possible impact of several types of crystalline defects on the typical Weibull failure statistics of 4H-SiC MOS devices [12]. A device can be defined robust when the amount of charge flowed in the insulator before breakdown is larger than 1 C/cm$^2$. As can be seen, the oxide growth quality only affects the intrinsic devices population (high charge-to-breakdown). On the other hand, surface defects, like down-falls, carrots and step-bunching, produce the appearance of the extrinsic breakdown population (low charge-to-breakdown). However, the role of threading dislocations (threading screw dislocations (TSDs) and threading edge dislocations (TEDs)) as killer defects for 4H-SiC MOSFETs is still unclear [13,14].

Hence, a deeper understanding of the breakdown phenomenon in relation to the presence of threading dislocations is still required, to pave the way for the achievement of reliable devices.

**Experimental**

In this paper, 4H-SiC MOSFETs fabricated to withstand 650V were used. The electrical results acquired on these devices after high temperature reverse bias (HTRB) test at 140°C and $V_{DS}$=600 V for $10^7$ s (three months) are discussed, and correlated with nanoscale characterization analyses (transmission electron microscopy (TEM), conductive atomic force microscopy (C-AFM) and scanning capacitance microscopy (SCM)).

A set of several hundreds of discrete transistors was stressed in HTRB configuration and each gate current was collected. An Infinity 1000 Qualitau Equipment was used for the HTRB. Such equipment is able to collect simultaneously the current flowing in each device.

The modification occurred during the HTRB stress at the SiO$_2$/4H-SiC interface was investigated by means of capacitance vs voltage (C-V) characterization of the failed and reference devices using a Keysight B1505.

To clarify the role of the TD on the dielectric breakdown, a nanoscale electrical characterization has been carried out to determine the electron properties of the material in the surrounding region of the TD. In particular, scanning probe microscopy by means of C-AFM and SCM were employed on the bare surface of the delayered power MOSFET on the breakdown region, using a DI3100 AFM by Veeco equipped with a Nanoscope V controller. Transmission electron microscopy (TEM) were performed using a Hitachi HD2300 STEM operated at 200 kV.

**Results and discussion**

Fig. 2a shows the gate current of two failed devices test at 140°C and $V_{DS}$=600 V for $10^7$ s. For one device the current was limited to $10^{-7}$ A (Soft breakdown Soft-BD device), while the second was not limited (Hard breakdown Hard-BD device). In the mentioned HTRB configuration, the maximum of the electric field in the insulator is about 4.6 MV/cm. As can be seen in Fig. 2b, the $I_D$-$V_D$ characteristics of the two failed devices are compared with those of a device that survived to the HTRB test (Reference Ref-D device). While the device Soft-BD shows a degradation of the output characteristics compared to Ref-D, the Hard-BD does not show the typical working device behavior, since no gate modulation on the drain conduction is observed (no $I_D$-$V_D$ is measured at high $V_G$). Analogously, by repeating three times the $I_D$-$V_G$ characteristics (Fig. 2c) on each device, it is possible to notice a similar behavior of Ref-D and Soft-BD (with no shift of the characteristics). On the other hand, Hard-BD shows a normally-on behavior and a shift of the transfer characteristics, thus indicating the occurrence of a charge trapping. Furthermore, the gate current (Fig. 2d) collected with a similar procedure shows an increase at $V_G$>2V for Soft-BD compared to Ref-D, thus indicating a partial degradation of the insulator due to an enhanced conduction. The Hard-BD shows a significant gate current enhancement and the shift occurring by repeating the measurements confirms the occurrence of charge trapping.

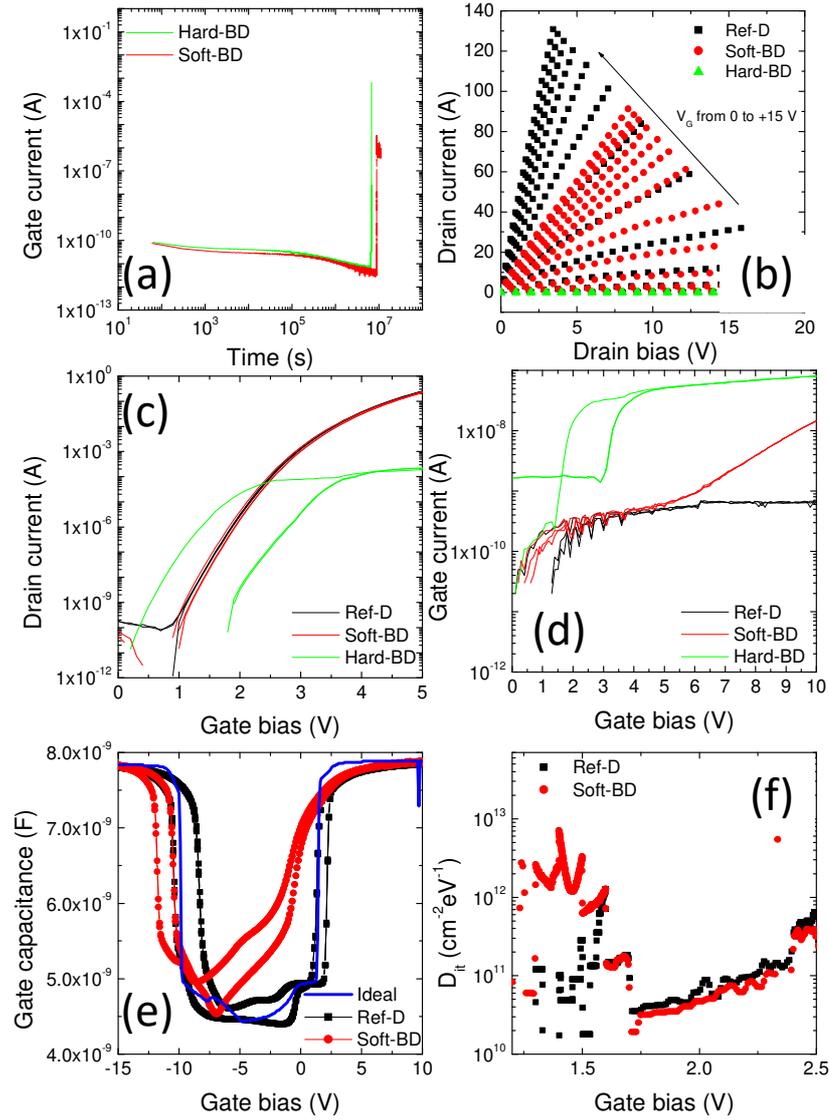

Fig. 2: (a) $I_G$-t in HTRB (b) $I_D$-$V_D$ (c) $I_D$-$V_G$ (d) $I_D$-$V_G$ (e) C-V and (f) $D_{it}$-$V_G$

Furthermore, the C-V curves collected at 1 kHz on the Ref-D and Soft-BD were compared with the ideal curve calculated with SILVACO tools (Fig. 2e). The C-V curve of the Ref-D is in good agreement with the ideal C-V curve, except for the hysteresis due to the charge/discharge of pre-existing interface states [15]. On the other hand, the C-V curve of Soft-BD is stretched (Fig. 2e), due to the presence of interface states. The interface states density ($D_{it}$) vs gate capacitance was determined from the curves stretch out and is reported in Fig. 2f. In power MOSFET, the gate contact and the gate insulator are fabricated covering simultaneously a p-type (body) and an n-type region (JFET). Hence, for a given gate bias value the energy levels in the body and in the JFET region are not identical. Hence, the $D_{it}$ profiling shown in Fig. 2f is not converted as a function of the energy position but it is kept as a function of the gate bias. Nevertheless, the direct comparison between identical devices can give useful information on the creation of interface states. The comparison of the $D_{it}$ profiles of the Ref-D and Soft-BD reveals that the $D_{it}$ near the 4H-SiC conduction band edge ($V_G$=2.5V) is about $10^{12} cm^{-2} eV^{-1}$ in both cases. On the other hand, in the case of Soft-BD an increase of the $D_{it}$ at mid-gap is observed, thus indicating the creation of these additional states during HTRB failure. Hence, it can be argued that these mid-gap states created during the HTRB can act as donor like introducing an effective positive charge in the MOS system. This can contribute with the progressive reduction of the gate current observed in Fig. 2a together with and hole-trapping in the insulator.

To get further insights on the dielectric breakdown phenomenon, a structural and electrical characterization at the nanoscale has been carried out on the failed devices. The location of the breakdown was found after the HTRB test by electron emission microscopy. Then, the devices were delayered exposing the bare 4H-SiC surface. The crystalline quality of the semiconductor material in proximity of the identified breakdown spot was investigated on the delayered devices by TEM analyses. TEM in cross section allowed to detect the presence of a threading dislocation (TD) running along the epitaxial layer from the substrate up to the surface. The TEM dual beam investigation ([11-20] and [0002]) shown in Figs. 3a and 3b allowed to establish the mixed nature of the dislocation. In particular, the dislocation shows up with the [11-20] spot (edge) and with the [0002] spot (screw). In fact, it was unclear whether threading dislocations (screw and edge) behave as killer defects in 4H-SiC MOSFETs. Threading dislocations have been suspected [13,16] and absolved [14,17] of inducing early device breakdown.

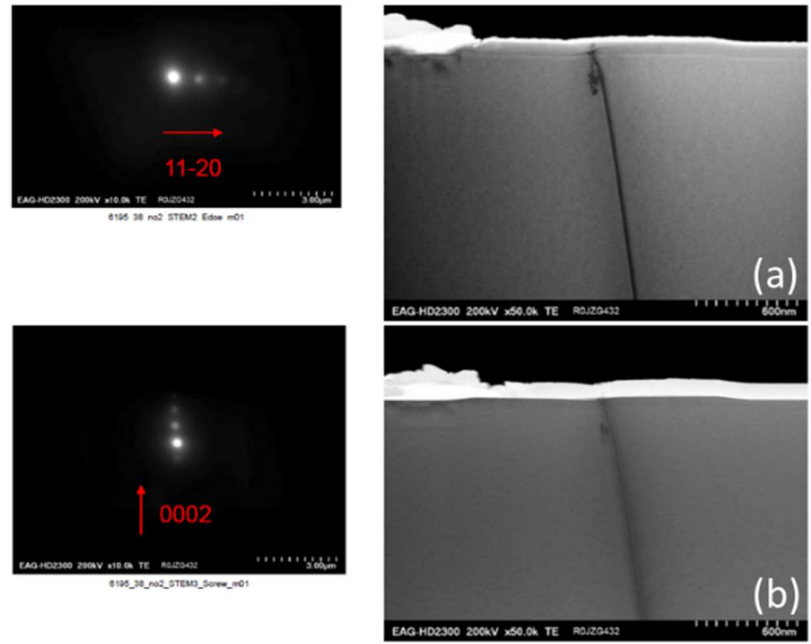

Fig. 3: (a) Cross section TEM in 2-beams configuration [11-20] (b) Cross section TEM in 2-beams configuration [0002].

AFM (Fig. 4a) performed in the JFET region, where the dislocation reaches the surface, allowed to visualize an isosceles triangle with the vertex in the [11-20] direction, about 25 nm deeper than the surface. Moreover, the C-AFM current map (Fig. 4b) showed a homogeneous conductivity on the surface. Noteworthy, the current is at least two orders of magnitude larger on the isosceles triangle vertex, i.e., where the TD is reaching the surface.

SCM collected simultaneously the impedance variation both in amplitude (Fig. 4c) and phase (Fig. 4d). Generally, the SCM signal amplitude is related to the net active dopants concentration in the region underneath the tip, while the phase signal is sensitive to the type carriers in the region underneath the tip. In the present case, the amplitude is not affected by the presence of the triangular defect (Fig. 4c), thus indicating no significant doping variation in the defect region. On the other hand, the SCM phase (Fig. 4d) varies within in the triangular defect region. In general, SCM phase variation suggests a variation of minority carrier concentration (holes) occurred in the investigated (JFET) region. The SCM experimental result agrees with the theoretical calculation provided by *Łazewski et al* [18] that employed density functional theory calculations of the energy levels and the effective band gap of the TD. They have found some localised broad states inside the TD, producing an effective reduction of the semiconductor band gap. This band gap shrinking occurs preferentially pulling up the valence band [18]. Hence, the variation in the valence band edge produces a variation in the generation/recombination of the minority carries in the JFET region where the TD is located and the breakdown occurred. Thus, the holes concentration in the JFET region can be noticeable increased.

The local increase of the holes concentration near the TD can be a source for hot holes injection into the insulator. This suggests that the local variation of the semiconductor electronic properties in the presence of the TD in the JFET region act as an accelerator for the insulator breakdown kinetics.

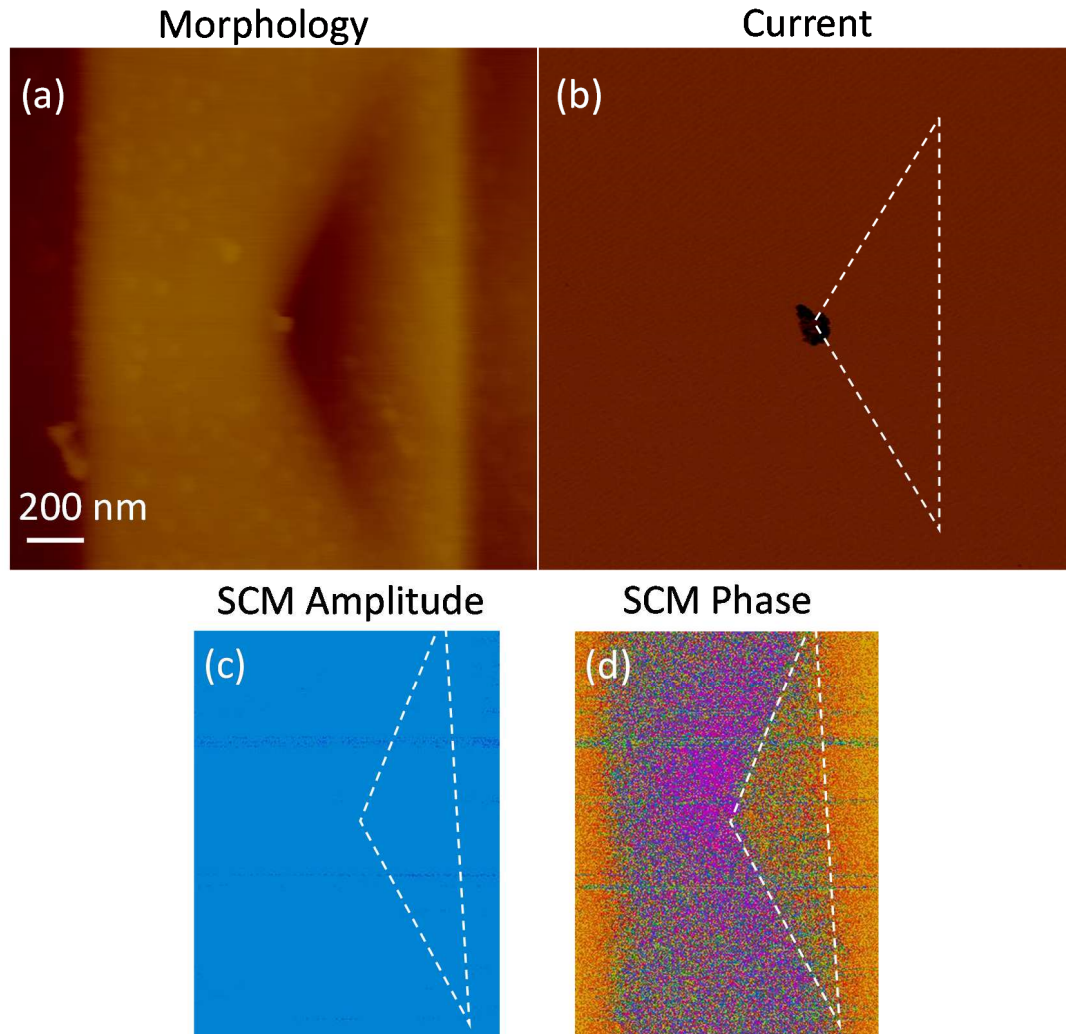

Fig. 4: (a) 4H-SiC AFM morphology in the JFET region (b) C-AFM Current map showing the enhanced conduction in the TD. (c) SCM amplitude showing the homogeneous doping concentration in the triangular defect region. (d) SCM phase showing variation in the triangular defect region.

**Conclusion**

In this work, a nanoscale electrical characterization by CAFM and SCM allowed us to explain the role of threading dislocations on the dielectric breakdown of 4H-SiC MOSFETs subjected to HTRB. In particular, a variation of the minority carries compatible with the semiconductor band gap narrowing was observed in the threading dislocations running through the 4H-SiC epitaxial layer. The increase of the minority carriers (holes) in the JFET region of the MOSFET enhances the holes injection into the insulator, thus accelerating the dielectric breakdown in correspondence of the threading dislocation.

## Acknowledgements

This work was carried out in the framework of the ECSEL JU project REACTION (first and euRopEAn siC eigTh Inches pilOt liNe), grant agreement no. 783158.

## References

[1] T. Kimoto, Jpn, J. Appl. Phys. 54, (2015) 040103
[2] P. Fiorenza, F. Giannazzo, F. Roccaforte, Energies, 12, (2019)2310
[3] H. Yano, et al ; IEEE Trans Elect Dev. 62, (2015) 324
[4] A. J. Lelis, R. Green, D. B. Habersat, M. El, IEEE Trans. Electron Devices 62, (2015) 316.
[5] P. Fiorenza, V. Raineri, Appl. Phys. Lett. 88, (2006) 212112
[6] J. H. Stathis, J. Appl. Phys. 86, (1999) *5757*
[7] P. Fiorenza, R. Lo Nigro, V. Raineri, D. Salinas, Materials Science Forum 556-557*, (2007) 501-504*
[8] E. T. Ogawa, J. Kim, G. S. Haase, H. C. Mogul, J. W. McPherson, "Leakage, breakdown and TDDB characteristics of porous low-K silicabased interconnect dielectrics," in Proc. Int. Rel. Phys. Symp., (2003), pp. 166–172
[9] E. Van Brunt, D. J. Lichtenwalner, R. Leonard, A. Burk, S. Sabri, B. Hull, S. Allen, J. W. Palmour, 2017 29th International Symposium on Power Semiconductor Devices and IC's (ISPSD): 28 May-1 June 2017 DOI: 10.23919/ISPSD.2017.7988907
[10] J. Eriksson, F. Roccaforte, P. Fiorenza, M.-H. Weng, F. Giannazzo, J. Lorenzzi, N. Jegenyes, G. Ferro, V. Raineri; J. Appl. Phys. 109, (2011) 013707
[11] O. Ishiyama, K. Yamada, H. Sako, K. Tamura, M. Kitabatake, J. Senzaki, H. Matsuhata; Jpn. J. Appl. Phys. 53, (2014) 04EP15
[12] M. Gurfinkel, J. C. Horst, J. S. Suehle, J. B. Bernstein, Y. Shapira, K. S. Matocha, G. Dunne, R. A. Beaupre; IEEE Trans Dev Mater Rel, 8, (2008) 635
[13] J. Senzaki, A. Shimozato, M. Okamoto, K. Kojima, K. Fukuda, H Okumura, K. Arai, Jpn. J. Appl. Phys. 48, (2009) 081404
[14] K. Matocha, G. Dunne, S. Soloviev, R. Beaupre, IEEE Trans. Electron Devices, 55, (2008) *1830–1834*
[15] P. Fiorenza et al., J.Vac. Sci. Tech. B 35, (2017)01A101
[16] J. Senzaki, K. Kojima, T. Kato, A. Shimozato, K. Fukuda; Appl. Phys. Lett., 89, (2006) 022909
[17] T. Watanabe, S. Hino, T. Iwamatsu, S. Tomohisa, S. Yamakawa, IEEE Trans Dev Mater Rel., 17, (2017) 163
[18] J. Łazewski, P. T. Jochym, P. Piekarz, M. Sternik, K. Parlinski, J. Cholewinski, P. Dłuzewski, S. Krukowski; J Mater Sci 54 (2019) 10737–10745